\documentclass[amsmath,amssymb,aps,prl,superscriptaddress,floatfix, preprint, %twocolumn %,reprint%
]{revtex4-2}

\usepackage{graphicx}% Include figure files
\usepackage{dcolumn}% Align table columns on decimal point
\usepackage{bm}
\usepackage[usenames]{color}
\usepackage{comment}
\usepackage{wasysym}
\usepackage{hyperref} 
\usepackage{physics}
\usepackage{bookmark}
\usepackage{mathtools}
\usepackage{subfigure}
\usepackage{longtable}
\usepackage{makecell}% make cells in tables
\usepackage{floatrow}% Table Caption position control
\usepackage[T1]{fontenc}%for some references
\usepackage[usenames]{color}
\usepackage[normalem]{ulem}
\usepackage{graphicx}
\usepackage{floatrow}

\newfloatcommand{figurebox}{figure}[\nocapbeside][\dimexpr(\textwidth-\columnsep)/2\relax]
\newfloatcommand{tablebox}{table}[\nocapbeside][\dimexpr(\textwidth-\columnsep)/2\relax]
\def\kk{{\bm k}}

\begin{document}

\title{Quantum metric-induced oscillations in nearly dispersionless flat bands}

\author{Hui Zeng}
\affiliation{State Key Laboratory of Low-Dimensional Quantum Physics, Department of Physics, Tsinghua University, Beijing 100084, China}
\affiliation{School of Physics, Peking University, Beijing 100871, China}
\author{Zijian Zhou}
\affiliation{State Key Laboratory of Low-Dimensional Quantum Physics, Department of Physics, Tsinghua University, Beijing 100084, China}

\author{Wenhui Duan}
\affiliation{State Key Laboratory of Low-Dimensional Quantum Physics, Department of Physics, Tsinghua University, Beijing 100084, China}
\affiliation{Institute for Advanced Study, Tsinghua University, Beijing 100084, China}
\affiliation{Frontier Science Center for Quantum Information, Beijing 100084, China}
\affiliation{Collaborative Innovation Center of Quantum Matter, Beijing 100871, China}

\author{Huaqing Huang}
\email[Corresponding author: ]{huaqing.huang@pku.edu.cn}
\affiliation{School of Physics, Peking University, Beijing 100871, China}
\affiliation{Collaborative Innovation Center of Quantum Matter, Beijing 100871, China}
\affiliation{Center for High Energy Physics, Peking University, Beijing 100871, China}

\date{\today}

% abstract
\begin{abstract}
The transport of Bloch electrons under strong fields is traditionally understood through two mechanisms: intraband Bloch oscillations and interband Zener tunneling. Here we propose a new oscillation mechanism induced by the interband quantum metric, which would significantly affect the electron dynamics under strong fields. By considering the multiband dynamics to the second order of the density matrix, we reveal that quantum metric-induced oscillations (QMO) persist regardless of band dispersion, even in exactly dispersionless flat bands. The resultant drift current can reach a magnitude comparable to the Bloch oscillation-induced drift current in systems where interband tunneling is negligible. Notably, the {QMO}-induced drift current increases linearly with electric field strength under the constraints of time-reversal or spatial-inversion symmetry, emerging as the primary delocalized current. We further show that both one-dimensional and two-dimensional superlattices are potential platforms for investigating QMO.
\end{abstract}

\maketitle
%\tableofcontents

\textit{Introduction.}---The strong-field dynamics of Bloch electrons has been an important area of research in condensed matter physics for decades. Among the well-known phenomena are Bloch oscillations (BO) \cite{bloch1928quantum}, arising from band dispersion, and interband Zener tunneling (ZT) \cite{zener1934theory}, particularly near anti-crossing points. While oscillations reflect the intrinsic periodicity of the system, interband tunneling disrupts these oscillations. {BO induces Wannier-Stark localization, with a steady drift current that decreases as the electric field increases, while ZT causes delocalization, resulting in a steady current that increases with the field.}
Extensive studies have explored these phenomena in one-dimensional (1D) superlattices, from strong-field transport in solids \cite{LEO1992943, PhysRevB.46.7252, PhysRevLett.70.3319, PhysRevLett.64.52, PhysRevLett.64.3167, PhysRevB.51.17275, PhysRevB.52.5105} to optical \cite{PhysRevLett.96.023901, PhysRevLett.102.076802, PhysRevLett.103.123601, PhysRevLett.91.263902, PhysRevLett.83.4756} and cold-atom \cite{PhysRevLett.87.140402, PhysRevLett.76.4508} simulations. 
%Extensive investigations into these phenomena have been conducted in one-dimensional (1D) superlattices, spanning studies on strong-field transport in solids \cite{LEO1992943, PhysRevB.46.7252, PhysRevLett.70.3319, PhysRevLett.64.52, PhysRevLett.64.3167, PhysRevB.51.17275, PhysRevB.52.5105} to simulations using optical \cite{PhysRevLett.96.023901, PhysRevLett.102.076802, PhysRevLett.103.123601, PhysRevLett.91.263902, PhysRevLett.83.4756} and cold-atom \cite{PhysRevLett.87.140402, PhysRevLett.76.4508} setups.
Recent breakthroughs in fabricating two-dimensional (2D) moir\'e superlattices featuring flat minibands \cite{lau2022reproducibility} have reignited interest in strong-field transport \cite{Bloch_in_Moire, PhysRevB.105.L241408, PhysRevB.107.165422}.
Furthermore, oscillations arising from the anomalous velocity induced by Berry curvature (i.e., Berry curvature-induced oscillations, BCO) \cite{Quantum_Geometric_Oscillations, Oscillation_Optical_Effects} has emerged as another significant mechanism influencing strong-field responses in 2D flat-band solids. This highlights the non-negligible role of quantum geometry in understanding the behavior of materials under strong fields.

Non-interacting, nearly dispersionless bands without anti-crossing points, referred to as flat bands in this work,  are commonly considered ideal platforms for observing intraband oscillations, i.e., BO \cite{Ivo_Souza_PRB, Bloch_in_Moire} and BCO \cite{Quantum_Geometric_Oscillations, Oscillation_Optical_Effects}. This is because interband ZT typically occurs near anti-crossing points where energy bands approach one another \cite{Quantum_Geometric_Oscillations, Ivo_Souza_PRB, Zener_Breakdown_PhysRevLett.86.1307, PhysRevB.107.165422, PhysRevLett.116.245301, Kitamura2019NonreciprocalLT, L-Z_Green_RPB_2020}.
%However, previous experiments showed that the standard BO-ZT picture is insufficient to understand the significant delocalized current observed in flat bands, and the interband coherence could play an important role \cite{PhysRevLett.64.3167}. In this work we demonstrate that such interband coherence is induced by quantum metric and can significantly affect the strong-field transport, emphasizing the necessity for a comprehensive exploration of interband quantum geometric effects.
%However, previous experiments have observed significant delocalized current in flat bands where tunneling is negligible \cite{PhysRevLett.64.3167}, indicating the need for a new mechanism beyond the conventional BO-ZT framework. In this work, we demonstrate that interband coherence driven by the quantum metric can strongly affect strong-field transport, emerging as a key source of delocalized current in flat bands, and highlighting the need for a deeper exploration of interband quantum geometric effects.
However, previous experiments have detected substantial delocalized current in GaAs/Al$_x$Ga$_{1-x}$As superlattices, a flat-band system where tunneling is negligible \cite{PhysRevLett.64.3167}, suggesting the need for a mechanism beyond the conventional BO-ZT framework. In this work, we demonstrate that interband coherence driven by the quantum metric can significantly affect strong-field transport, emerging as a key contributor to delocalized currents in flat bands. This highlights the necessity of a deeper investigation into interband quantum geometric effects.

In this Letter, we reveal a new type of oscillation driven by the interband quantum metric which characterizes the distance of Bloch states between different bands. We show that the quantum metric-induced oscillations (QMO) occur even in exactly flat bands where conventional BO vanishes.
Remarkably, the {delocalized} drift current from QMO exhibits a linear growth asymptotic behavior in the strong-field regime, which is distinct from the BO-induced {localized} drift current decaying as $E^{-1}$ or the BCO-induced drift current saturating to an $E$-independent value.  
Notably, the QMO-induced drift current in flat-band systems can reach a magnitude comparable to the BO-induced drift current when the system is far from ZT-induced breakdown, thus representing a significant mechanism beyond interband tunneling that resists the BO-induced localization. Furthermore, we show that both 1D and 2D superlattices are potential platforms for investigating the QMO.

\textit{Recursive formula for strong-field transports.}---We start by establishing the recursive formula of the density matrix to address the multi-band-coupled strong field problems{, which requires a derivation beyond the standard weak-field expansion \cite{PhysRevLett.112.166601, PhysRevLett.127.277201, PhysRevLett.127.277202, PhysRevLett.131.056401, gao2023quantum, kaplan2023unification, lahiri2023intrinsic, wang2023quantum}}. To study the electronic response properties under an external electric field, we begin with the master equation in the relaxation-time approximation \footnote{\label{fn}See Supplemental Material at http://link.aps.org/supplemental/xxx, for more details about the approximations, derivations of the drift \& oscillatory currents, numerical calculations, and possible experimental realizations. Refs.~\cite{QianNiu_Relaxation, RTA1_PhysRevB.90.195429, RTA2_PhysRevB.91.155422, BeyondRTA1_PhysRevX.5.041050, BeyondRTA2_PhysRevB.91.184301, lau2022reproducibility, Bloch_in_Moire, PhysRevB.105.L241408, PhysRevB.107.165422,Quantum_Geometric_Oscillations, Oscillation_Optical_Effects,kaplan2023unification, lahiri2023intrinsic, PhysRevLett.131.056401, gao2023quantum, wang2023quantum, PhysRevX.11.011001, PhysRevLett.115.216806, PhysRevLett.112.166601, PhysRevLett.127.277202, PhysRevLett.127.277201, Kitamura2019NonreciprocalLT, L-Z_Green_RPB_2020, PhysRevB.104.085114, PhysRevLett.96.023901, PhysRevLett.102.076802, PhysRevLett.103.123601, PhysRevLett.91.263902, PhysRevLett.83.4756,PhysRevLett.87.140402, PhysRevLett.76.4508,guo2021observation,LEO1992943, PhysRevB.46.7252, PhysRevLett.70.3319, PhysRevLett.64.52, PhysRevLett.64.3167, Angeli2020TTMD, cao2018unconventional, cao2018correlated, cai2023signatures, zeng2023thermodynamic, park2023observation, xu2023observation, kang2024evidence, lu2024fractional, yin2022topological, wang2023Kagome, Du2021NonlinearHE, PhysRevLett.131.076601, Essay_Quantum_Geometry, PhysRevResearch.1.032019, PhysRevResearch.5.L032016, PhysRevB.97.201117, PhysRevLett.122.210401, PhysRevLett.122.227402,PhysRevLett.131.156901} are included.}:
%\cite{QianNiu_Relaxation, RTA1_PhysRevB.90.195429, RTA2_PhysRevB.91.155422, BeyondRTA1_PhysRevX.5.041050, BeyondRTA2_PhysRevB.91.184301, lau2022reproducibility, Bloch_in_Moire, PhysRevB.105.L241408, PhysRevB.107.165422,Quantum_Geometric_Oscillations, Oscillation_Optical_Effects,kaplan2023unification, lahiri2023intrinsic, PhysRevLett.131.056401, gao2023quantum, wang2023quantum, PhysRevX.11.011001, PhysRevLett.115.216806, PhysRevLett.112.166601, PhysRevLett.127.277202, PhysRevLett.127.277201, Kitamura2019NonreciprocalLT, L-Z_Green_RPB_2020, PhysRevB.104.085114, PhysRevLett.96.023901, PhysRevLett.102.076802, PhysRevLett.103.123601, PhysRevLett.91.263902, PhysRevLett.83.4756,PhysRevLett.87.140402, PhysRevLett.76.4508,guo2021observation,LEO1992943, PhysRevB.46.7252, PhysRevLett.70.3319, PhysRevLett.64.52, PhysRevLett.64.3167, Angeli2020TTMD}
\begin{equation}\label{eq1: Master Equation}
\begin{aligned}
    i \hbar \partial_t \rho = \left[H_0 + H_1,\rho\right] - i(\rho -\rho^0)\hbar / \tau,
\end{aligned}
\end{equation}
where $\rho(\bm{k},t)$ is the density matrix operator in momentum space, and hereafter we omit the explicit dependence of $\bm{k}$ for brevity, unless otherwise specified. $\rho^0 = f_n^0 \delta_{nm}$ is the equilibrium Fermi-Dirac distribution{, and we assume zero temperature unless otherwise specified}. $H_0$ is the Hamiltonian without an external field, $\tau$ is the relaxation time, and $H_1=e\bm{E}\cdot {\bm{r}}$ captures the coupling between the external electric field and electron. The position operator in the Bloch basis is represented as $[\bm{r}_{\bm{k}}]_{nm} = i\partial_{\bm{k}}\delta_{nm} + \mathcal{A}_{nm}$, where $\mathcal{A}$ is the Berry connection. 

It should be noted that four energy scales appear in Eq.~\eqref{eq1: Master Equation}: the characteristic frequency $\hbar \omega$ ($i \hbar \partial_t$) of the density matrix evolution; the energy gap $\epsilon_{nm}=\epsilon_n - \epsilon_m $ from the eigenvalues of $H_0$; the external field strength characterized by $e{E}{a}$ from $H_1$ with ${a}$ the lattice {constant}; and the relaxation process characterized by $\hbar/\tau$. 
In this work, we consider the large-gap condition, i.e., $\epsilon_{nm} \gg \max\{\hbar \omega, e{E}{a}, \hbar/\tau\}$, which includes two more specific cases.
One is characterized by a weak field ($\epsilon_{nm} \gg \hbar/\tau \gg e{E}{a}$) where various weak-field response properties can be obtained \cite{RevModPhys.82.1539, PhysRevLett.115.216806, Ma2018ObservationOT, Kang2019NonlinearAH, PhysRevLett.112.166601, PhysRevLett.127.277201, PhysRevLett.127.277202, PhysRevLett.131.056401, gao2023quantum, wang2023quantum, kaplan2023unification, lahiri2023intrinsic, PhysRevLett.131.056401, gao2023quantum, wang2023quantum}.  
The other condition pertains to a strong field under the large-gap presupposition %/premise/precondition/criterion %
($\epsilon_{nm} \gg e{E}{a} \gg \hbar/\tau$). This refined energy scale condition, which is our main concern, would dedicate fascinating situations where $\hbar \omega \neq 0$ leads to oscillatory currents, while $\hbar \omega = 0$ corresponds to drift currents, as we discussed later.

Considering the time scale of BO ($\omega \sim eEa/\hbar$) and steady transports, we expand Eq.~\eqref{eq1: Master Equation} under the large-gap condition. This leads us to a recursive formula for the off-diagonal and diagonal terms of the density matrix $\rho(t)$:
\begin{equation}\label{eq2: Recursive nm}
\begin{aligned}
    \epsilon_{nm} \rho_{nm}^{(N)}(t) = & - e\bm{E}\cdot\left[\mathcal{A},\rho^{(N-1)}(t)\right]_{nm} \\
    &- i\left(e\bm{E}\cdot\partial_{\bm{k}}-\hbar \partial_{t}  -\hbar/\tau \right) \rho_{nm}^{(N-1)}(t),\\
\end{aligned}
\end{equation}
\begin{equation}\label{eq3: Recursive nn}
\begin{aligned}
    0 =    & - e\bm{E}\cdot\left[\mathcal{A},\rho^{(N)}(t)\right]_{nn} \\
    &     -i\left(e\bm{E}\cdot\partial_{\bm{k}}-\hbar \partial_{t} -\hbar/\tau \right)\rho_{nn}^{(N)}(t),\\
\end{aligned}
\end{equation}
where $\rho^{(N)}$ represents the $N$-th order term of $1/\epsilon_{nm}$, $\left[\mathcal{A},\rho^{(N)}(t)\right]_{nm} = \sum_{p} \left(\mathcal{A}_{np}\rho_{pm}^{(N)}-\rho_{np}^{(N)}\mathcal{A}_{pm}\right)$. These formulas apply to both weak-field and strong-field cases, as shown in Supplementary Materials (SM) \footnotemark[\value{footnote}]. 
Furthermore, we introduce the diagonal density matrix $\rho^{(0)}$, which satisfies the master equation Eq.~\eqref{eq1: Master Equation} at the zeroth order, as the starting point for the recursive process
\begin{equation}\label{eq4: Steady 0}
\begin{aligned}
    &\rho^{(0)}_{\text{s},{nm}} = \sum_{\bm{a}_{i}} \frac{f_{n,\bm{a}_i}^0 e^{i\bm{k}\cdot \bm{a}_{i}}} {1-i\tau e\bm{E}\cdot\bm{a}_{i}/\hbar}   \delta_{nm} , \quad t\gg \tau\\
\end{aligned}
\end{equation}
\begin{equation}\label{eq5: Oscillation nn 0}
\begin{aligned}
    %&\rho^{(0)}(t)=\rho_{nn,t=0}(\bm{k}+e\bm{E}t/\hbar)\delta_{nn} 
    &\rho^{(0)}_{{nm}}(\kk,t)=\rho_{nn}(\bm{k}+e\bm{E}t/\hbar,0)\delta_{nm},\quad t\ll \tau
\end{aligned}
\end{equation}
where $f^0_{n,\bm{a}_i}$ is $\bm{a}_i$ Fourier component of Fermi-Dirac distribution. $\rho^{(0)}_{\text{s}}$ is the zeroth order density matrix for the steady state after full relaxation %when system is fully relaxed
\cite{Quantum_Geometric_Oscillations},
and $\rho^{(0)}(\kk,t)$ represents the zeroth order density matrix for the non-steady oscillating state far from complete relaxation. {Note that under the large gap condition, the small quantity is $1/\epsilon_{nm}$  instead of $E$. $E$ is allowed to appear in $\rho^{(0)}$, capturing the non-perturbative effects of a strong electric field on the momentum-space electron distribution.}

Having obtained the density matrix, we can derive the electric current $\bm{J}=-e\operatorname{Tr} \left[ \rho \bm{v} \right]$, where the velocity operator can be expressed in the Bloch basis as $[\bm{v}]_{nm} = \frac{1}{\hbar} (\partial_{\bm{k}}\epsilon_n \delta_{nm} + i\epsilon_{nm}\mathcal{A}_{nm}) $. 
The $\omega = 0$ steady component of the \textit{zeroth-order} and \textit{first-order} density matrix respectively give the BO and BCO induced drift currents \cite{Quantum_Geometric_Oscillations} in 2D under the strong-field condition
\begin{equation}\label{eq6: Bloch}
\begin{aligned}
    \bm{J}_{\text{Bloch}} 
    = & -\frac{e}{\hbar} \int_{\rm{BZ}} [d\bm{k}] \sum_{n} \rho^{(0)}_{\text{s},nn} \nabla_{\bm{k}} \epsilon_{n}, \\
\end{aligned}
\end{equation}
\begin{equation}\label{eq7: BC}
\begin{aligned}
    \bm{J}_{\Omega}     
    = & -\frac{e}{\hbar} \int_{\rm{BZ}} [d\bm{k}] \sum_{n} \rho^{(0)}_{\text{s},nn} e \bm{E} \times \Omega_{n}, \\
\end{aligned}
\end{equation}
where the integral is over the 2D Brillouin zone (BZ) with $[d\bm{k}]\equiv d^2 k/(2\pi)^2$, Berry curvature $\Omega_{n} = 2\sum_m\Im(\mathcal{A}_{nm}^x\mathcal{A}_{mn}^y)$. Correspondingly, we can also obtain the BO and BCO oscillatory currents, which share similar expressions as Eqs.~\eqref{eq6: Bloch} and \eqref{eq7: BC} only with $\rho_s^{(0)}$ replaced by $\rho^{(0)}(t)$. These expressions of BO and BCO are consistent with those derived from the semiclassical wave packet dynamics for a single band \cite{Quantum_Geometric_Oscillations}. Moreover, our recursive formula allows us to comprehensively address both intraband and interband effects simultaneously.
%These expressions of BO and BCO are consistent with recent work \cite{Quantum_Geometric_Oscillations}, which validates our derivations based on the recursive formula.

\textit{{QMO}-induced Drift current.}---Now, we  take a step further to consider the \textit{second-order} density matrix {of steady state}, which results in {drift current induced by} QMO ($\bm{J}_{g}$) and ZT ($\bm{J}_{\text{ZT}}$). 
Specifically, $\bm{J}_{g}$ and $\bm{J}_{\text{ZT}}$ are determined by the second-order off-diagonal and diagonal terms of the density matrix, derived from Eqs. \eqref{eq2: Recursive nm} and \eqref{eq3: Recursive nn}, respectively.
Although $\bm{J}_{\text{ZT}}$ is much smaller than $\bm{J}_{{g}}$ in flat-band systems, as illustrated later, here we introduce both for the sake of completeness, 
\begin{eqnarray}
%\begin{aligned}
    \bm{J}^{a}_{{g}}
    &=& -\frac{e^2 E^c}{\hbar} \int_{\rm{BZ}} [d\bm{k}] \sum_{n,m\neq n} \rho^{(0)}_{\text{s},nn} \frac{2{g}_{nm}^{ac}}{\epsilon_{nm}} \hbar/\tau \nonumber\\
    & &- \frac{e^3 {E}^b E^c}{\hbar} \int_{\rm{BZ}} [d\bm{k}]  \sum_{n,m\neq n}  \Big[\rho^{(0)}_{\text{s},nn} \frac{(\partial_a {g}_{nm}^{bc})}{\epsilon_{nm}} \Big], \label{eq8: JGG}\\
%\end{aligned}
%\begin{equation} \label{eq9: JGE}
    \bm{J}^a_{\text{ZT}} &=& -\frac{e}{\hbar} \int_{\rm{BZ}} [d\bm{k}] \sum_{n} [\rho_{nn}^{(2)} \nabla_{a} \epsilon_n], \label{eq9: JGE} \\
%\end{aligned}
%\begin{equation} \label{eq10: rho_nn^2}
    \rho_{nn}^{(2)} &=& \sum_{\bm{a}_i}  \frac{e^{i\bm{k}\cdot \bm{a}_i}}{e\bm{E}\cdot \bm{a}_i + i\hbar/\tau} \sum_{m\neq n} \Big[ e^2 E^d E^c \frac{i\hbar}{\tau}  2{g}_{nm}^{dc} \frac{\rho^{(0)}_{\text{s},nn}}{\epsilon_{nm}^2} \nonumber\\
    & &+ \frac{ie^3 E^d E^b E^c }{\epsilon_{nm}}   \Big((\partial_b {g}_{nm}^{dc}) \frac{\rho^{(0)}_{\text{s},nn}}{\epsilon_{nm}} + 2 {g}_{nm}^{dc}\partial_b(\frac{\rho^{(0)}_{\text{s},nn}}{\epsilon_{nm}})\Big) \nonumber\\
    & &- (n\leftrightarrow m)\Big]_{\bm{a}_i}, \label{eq10: rho_nn^2} 
\end{eqnarray}
where $a,b,c$ are Cartesian indices (Einstein summation convention assumed),
$\left[\cdot\right]_{\bm{a}_i}$ denotes the $\bm{a}_i$ Fourier component, and $g_{nm}^{ab} = \Re(\mathcal{A}_{nm}^a \mathcal{A}_{mn}^b)$ is the quantum metric. 
Unlike BCO-induced drift current $\bm{J}_{\Omega}$, which has only a transverse component and vanishes in 1D systems \cite{Quantum_Geometric_Oscillations}, $\bm{J}_{g}$ has both longitudinal and transverse components, making it detectable in both 1D and higher-dimensional systems.
%{Previous works have shown that quantum metric is crucial in nonlinear transport at weak fields}
%\cite{PhysRevLett.112.166601, PhysRevLett.127.277201, PhysRevLett.127.277202, PhysRevLett.131.056401, gao2023quantum, kaplan2023unification, lahiri2023intrinsic, wang2023quantum}
%{, our proposed QMO-induced drift current further clarify the importance of interband quantum geometric effects under both weak and strong field conditions.}

Notably, in systems with symmetry of time reversal ($\mathcal{T}$) or space inversion ($\mathcal{P}$), $\bm{J}_{{g}}$ is proportional to $E$ asymptotically, although Eq.~\eqref{eq8: JGG} appears to include $E^2$ terms.
This is because the fluctuations of $\rho^{(0)}_{\text{s},nn}$ are always suppressed by factor $1/(e\bm{E}\cdot \bm{a}_i)$, as shown in Eq.~\eqref{eq4: Steady 0}.
In the second term of $\bm{J}_{{g}}$ (Eq.~\eqref{eq8: JGG}), $\mathcal{T}$ or $\mathcal{P}$ symmetry constrains $g_{nm}$ and $\epsilon_{nm}$ to be even to $\bm{k}$, which in turn requires that only the $\bm{k}$-odd fluctuations of $\rho^{(0)}_{\text{s},nn}$ contribute to the integration. Therefore, both terms in Eq.~\eqref{eq8: JGG} are linear to $E$.
The same symmetry constraint also applies to $\bm{J}_{\text{ZT}}$, and an additional $1/(e\bm{E}\cdot \bm{a}_i) $ factor is introduced for $\rho^{(2)}_{\text{s},nn}$, as shown in Eq.~\eqref{eq10: rho_nn^2}. This leads to $\bm{J}_{\text{ZT}}$ also being proportional to $E$. More detailed discussions about the asymptotical behaviors under symmetry constraints are presented in SM \footnotemark[\value{footnote}].

\begin{figure}
    \centering
    \includegraphics[width=1.0\linewidth]{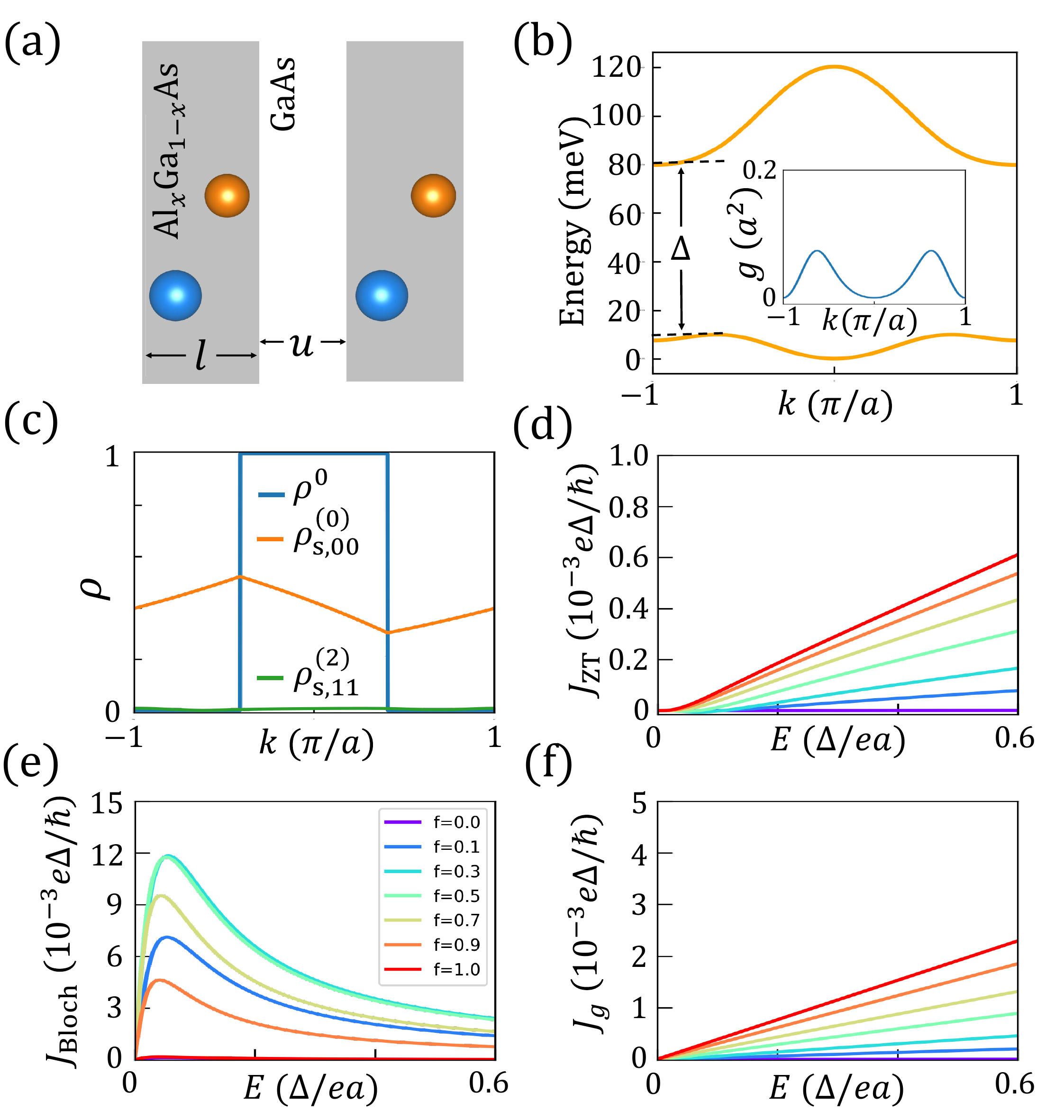}
    \caption{Drift currents in GaAs/Al$_{x}$Ga$_{1-x}$As superlattices. (a) Sketch of an effective 1D model with two orbitals for the two bottom conduction bands of the superlattice. (b) Band structure and quantum metric of the 1D model. (c) Equilibrium (Fermi-Dirac) filling distribution $\rho^0$ at band filling $f=0.4$, zeroth order distribution $\rho^{(0)}_{00}$ under a strong field of $E=\Delta/2ea$, and ZT-induced upper band occupation $\rho^{(2)}_{11} = -\rho^{(2)}_{00}$. (d-f) $J_{\text{ZT}}$, $J_{\text{Bloch}}$ and $J_{g}$ as functions of field strength $E$ for the {lower} band with different fillings [denoted as $f$ in (e)].}
    \label{fig: Drift_1D}
\end{figure}

We first illustrate the drift currents in a 1D GaAs/Al$_{x}$Ga$_{1-x}$As superlattice, which was usually used to study the BO \cite{LEO1992943, PhysRevB.46.7252, PhysRevLett.70.3319, PhysRevLett.64.52, PhysRevLett.64.3167, PhysRevB.51.17275, PhysRevB.52.5105}. We adopt a 1D effective model to describe the two bottom conduction minibands of the superlattice composed of alternating GaAs and Al$_{0.3}$Ga$_{0.7}$As layers with width $u=95$ \r{A} and $l=25$ \r{A} \cite{PhysRevB.52.5105}, as shown in Fig.~\ref{fig: Drift_1D}(a). The specifics of these calculations are detailed in SM \footnotemark[\value{footnote}]. Figure~\ref{fig: Drift_1D}(b) shows the calculated band structure and quantum metric, where the bandwidths are $W_{0,1}=10$ and $40$ meV and the energy gap between them is $\Delta=70$ meV. Taking a typical relaxation time $\tau \simeq 0.13$ ps, which satisfies the large-gap approximation (i.e., $\hbar/\tau = \Delta/14\ll\Delta$), we calculate the zeroth order {lower} band occupation $\rho^{(0)}_{\text{s},00}$ and ZT-induced conduction band occupation $\rho^{(2)}_{\text{s},11} = -\rho^{(2)}_{\text{s},00}$ in the presence of an external field of ${E} = \Delta/2ea$. As shown in Fig.~\ref{fig: Drift_1D}(c), the negligible $\rho^{(2)}_{\text{s},11} (\ll 1)$ ensures the system is far away from the breakdown, which indicates a small tunneling current ${J}_{\text{ZT}}$ $(\ll {J}_{\text{Bloch}},{J}_{{g}})$ [see Fig.~\ref{fig: Drift_1D}(d)]. 

Importantly, we find a considerable drift current ${J}_{{g}}$, which is comparable to ${J}_{\text{Bloch}}$, as shown in Figs.~\ref{fig: Drift_1D}(e,f).
Moreover, ${J}_{{g}}$ increases linearly under large electric fields, while ${J}_{\text{Bloch}}$ decays with the increasing $E$.
This indicates that in flat-band systems, contrary to the traditional view that ZT is the primary mechanism for delocalized current, $\bm{J}_{g}$ is actually the main source of such current.
Previous research in 1D superlattices has observed a transition from localized current with negative differential resistance to delocalized current, with ZT being negligible \cite{PhysRevLett.64.3167}. We suggest that $\bm{J}_{g}$ should be considered an important source of the delocalized current.
Additionally, unlike $J_{\text{Bloch}}$ which reaches the maximum in half-filled bands but vanishes in fully-filled bands, $J_{g}$ increases with the filling factor, highlighting its origin in interband coupling.

\begin{figure}
    \centering
    \includegraphics[width=1.0\linewidth]{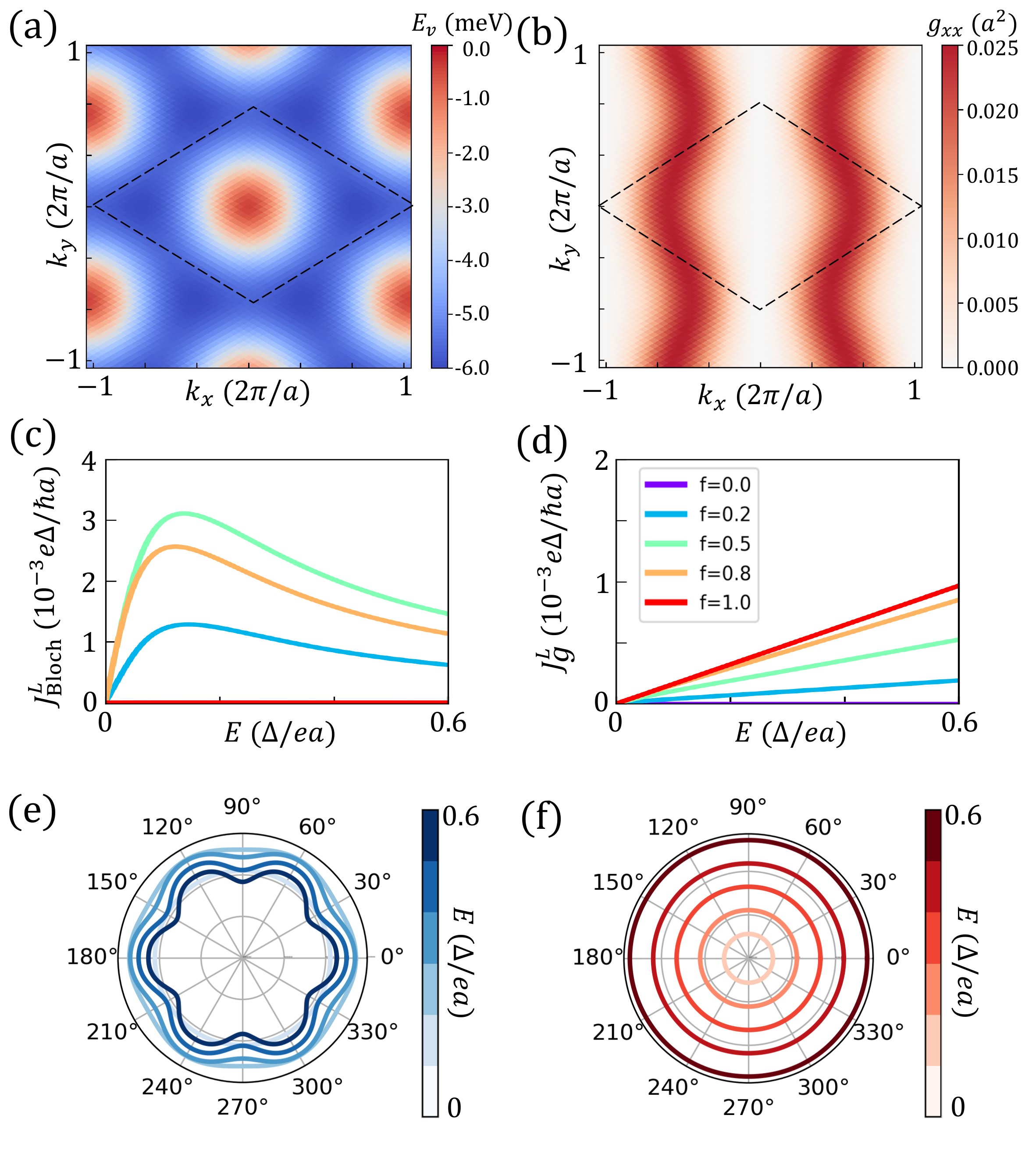}
    \caption{Bloch and quantum metric induced drift currents in 2D honeycomb lattice. (a) Valence band $E_v$ and (b) $g_{xx}$ component of quantum metric in the 2D honeycomb lattice model. (c,d) show the longitudinal component of $J_{\text{Bloch}}$ and $J_{g}$ varying with external electric field $E$ directed in $x$ when valence band has different fillings. (e,f) show how ${J}_{\text{Bloch}}$ and ${J}_{{g}}$ vary with electric field directions and field strengths, at a filling of 0.6.}
    \label{fig: Drift_2D}
\end{figure}

We further study the QMO-induced drift current $\bm{J}_{g}$ in a 2D honeycomb lattice model \footnotemark[\value{footnote}], which has been suggested as a good platform for studying $\bm{J}_{\text{Bloch}}$ \cite{Bloch_in_Moire} and $\bm{J}_{\Omega}$ \cite{Quantum_Geometric_Oscillations, Oscillation_Optical_Effects}. Here, we only present two dominant longitudinal currents ${J}^L_{\text{Bloch}}$ and ${J}^L_{{g}}$, and defer detailed discussion about small $\bm{J}_{\text{ZT}}$ and transverse currents to SM \footnotemark[\value{footnote}].
Figures~\ref{fig: Drift_2D}(a,b) display the valence band energy $E_{v}$ and the $g_{xx}$ component of the quantum metric, which are the origins of ${J}^L_{\text{Bloch}}$ and ${J}^L_{g}$, respectively, when the electric field $E$ is directed along the $x$-axis.
The nontrivial $E_{v}$ and $g_{xx}$ induce ${J}^L_{\text{Bloch}}$ and ${J}^L_{{g}}$, which exhibit inverse decay and linear increase with $E$, respectively, as shown in Figs.~\ref{fig: Drift_2D}(c,d).
The magnitudes of ${J}_{\text{Bloch}}$ and ${J}_{{g}}$ are comparable.
Moreover, ${J}_{\mathrm{Bloch}}$ shows a hexagonal dependence on the electric field direction, while ${J}_{{g}}$ exhibits an isotropic asymptotic behavior, as displayed in Figs.~\ref{fig: Drift_2D}(e,f). These features indicate that $J_g$ constitutes a significant delocalized current in 2D systems.
%with a Hamiltonian $ H = t \sum_{i=1,2,3} \cos(\bm{k}\cdot\bm{b}_i) \sigma_x+\sin(\bm{k}\cdot\bm{b}_i) \sigma_y + m \sigma_z + 2t' \sum_{i=1,2,3} \cos(\bm{k}\cdot\bm{a}_i) \sigma_0$, where $\bm{b}_{1,2,3}$($\bm{a}_{1,2,3}$) are the nearest (next-nearest) neighbor vectors, $\sigma$ is Pauli matrix for sublattice degree of freedom, $t$($t'$) is the hopping, and $m$ denotes a staggered AB-sublattice potential.

\textit{Quantum metric-induced oscillations.}---Next, we turn to time-dependent cases to study the quantum metric-induced oscillatory current. Specifically, we consider the time evolution of both the diagonal $\rho_{nn}^{(0)}(k,t)$ and off-diagonal $\rho_{nm}^{(0)}(k,t)$ density matrices from arbitrary initial states, which leads to oscillations at low and high frequencies (i.e., BO and $\epsilon_{nm}$ frequency), respectively.
Given that the recursive formula Eq.~\eqref{eq2: Recursive nm} holds for both steady ($\omega=0$) and oscillatory ($\omega \sim eEa/\hbar$) states, the QMO-induced oscillatory current ${J}_{{g}}(t)$ at BO frequency can be obtained by simply replacing $1/\tau$ and $\rho_{\text{s}}^{(0)}$ in Eq.~\eqref{eq8: JGG} with $i\omega$ and $\rho^{(0)}_{nn}(\omega)$. 
In particular, ${J}_{{g}}(t)$ in 1D systems can be expressed as \footnotemark[\value{footnote}],
\begin{equation}\label{eq11: Oscillation_Flat}
\begin{aligned}
   {J}_{{g},\text{BO-freq}}(t) = & -\frac{e^2 E}{\hbar} \int \frac{d {k}}{2\pi} \sum_{n,m\neq n} \partial_t \rho_{nn}^{(0)}(k,t) \frac{{g}_{nm} }{\epsilon_{nm}}.
   %{J}_{{g}}(\omega)    = & -i\omega {e^2 E} \int \frac{d {k}}{2\pi} \sum_{n,m\neq n} \rho_{nn}^{(0)}(\omega) \frac{{g}_{nm} }{\epsilon_{nm}}
\end{aligned}
\end{equation}
%which is the Bloch-frequency component of QMO with linear-to-$E$ amplitude.
Alternatively, a high-frequency oscillatory current arises from the off-diagonal density matrices, whose zeroth order term in flat bands is $\rho_{nm}^{(0)}({k},t)=\rho_{nm}({k}+e{E}t/\hbar,0)e^{-i\epsilon_{nm}t/\hbar}$ and higher-order low-frequency modulation can be derived with recursive formulas under the large-gap approximation (see SM \footnotemark[\value{footnote}] for detailed discussions).
Specifically, for a 1D state starting with density matrix with off-diagonal component $\rho_{nm}(k,0)$, its time-evolution is characterized by $\rho_{nm}(k,t)=e^{-i\epsilon_{nm}t/\hbar}C_{nm}(k,t)\mathcal{A}_{nm}(k)$, where $C_{nm}(k,t)$ determined by $\rho_{nm}({k}+e{E}t/\hbar,0)=C_{nm}(k,t)\mathcal{A}_{nm}(k)$ oscillates with a period equal to that of Bloch oscillations.
The $\epsilon_{nm}$-frequency component of the QMO-induced oscillatory current ${J}_{{g}}(t)$ is given by
\begin{equation}\label{eq12: Modulation}
\begin{aligned}
   {J}_{{g},\epsilon\text{-freq}}(t) = &  \frac{e}{\hbar} \int \frac{d {k}}{2\pi} \sum_{n,m\neq n} \operatorname{Im} \left(e^{-i \epsilon_{nm} t/\hbar}C_{nm}(k,t)\right) \epsilon_{nm} {g}_{nm}.   \\
   %{J}_{{g}}(\omega)    = & -i\omega {e^2 E} \int \frac{d {k}}{2\pi} \sum_{n,m\neq n} \rho_{nn}^{(0)}(\omega) \frac{{g}_{nm} }{\epsilon_{nm}}
\end{aligned}
\end{equation}
Although it does not generate a net current on the BO time scale, as the high-frequency current is averaged out, it can dominate the breathing mode produced by the evolution of wave packets on the lattice.

\begin{figure}
    \centering
    \includegraphics[width=1.0\linewidth]{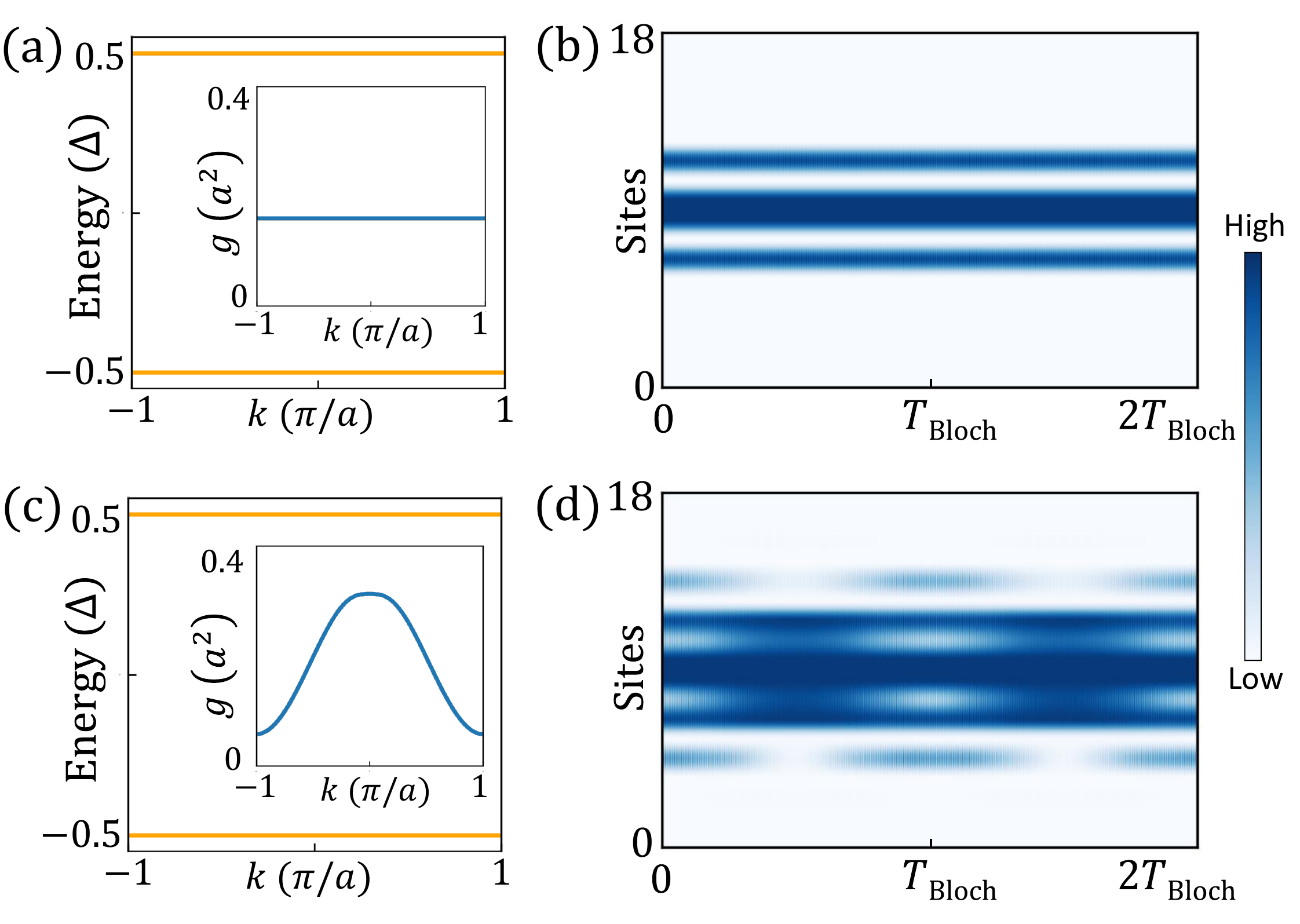}
    \caption {Quantum metric-induced oscillations lead to breathing modes of probability density in 1D lattice model with 18 sites. (a,b) Model with exactly flat bands and constant quantum metric $g$ do not exhibit any oscillations. (c,d) When a non-constant quantum metric is introduced into the exactly flat bands, QMO-induced breathing mode emerges, and probability density evolves periodically in time. {Two orbitals are shown on two sites separately.}}
    \label{fig: Oscillations}
\end{figure}

Notably, $J_{g}$ occurs even in exactly flat bands when $g$ varies in momentum space, but vanishes if both $\epsilon_{nm}$ and $g_{nm}$ are constant. This is because both $\rho^{(0)}_{nn}(k,t)=\rho_{nn}({k}+e{E}t/\hbar,0)$ and $\rho_{nm}(k,t) 
= \rho_{nm}(k+eEt/\hbar,0)$ are just shifting within the Brillouin zone, and therefore, when $\epsilon_{nm}$ and $g_{nm}$ are constant, the integrals in Eqs.~\eqref{eq11: Oscillation_Flat} and \eqref{eq12: Modulation} do not change {in BO time scale}. However, when $g$ varies, it imparts a non-constant weight to the integrals, causing the integral results periodically evolve at the Bloch frequency. In addition, the amplitudes of ${J}_{{g},\text{BO-freq}}(t)$ and ${J}_{{g},\epsilon\text{-freq}}(t)$ do not decay with increasing electric field \footnotemark[\value{footnote}]. This characteristic distinguishes them from BO and provides potential probes of the quantum metric in periodic systems.

1D lattices are excellent platforms for studying BO \cite{PhysRevLett.96.023901, PhysRevLett.102.076802, PhysRevLett.103.123601, PhysRevLett.91.263902, PhysRevLett.83.4756, PhysRevLett.87.140402, PhysRevLett.76.4508} and can also be used to investigate QMO. The periodic evolution of the wavepacket probability distribution over the lattice, known as the breathing mode, visualizes the oscillatory behavior. To directly illustrate QMO, we {numerically simulate the time-dependent evolution of a wave packet in} a 1D two-band model with exactly flat bands, where BO vanishes due to the absence of band dispersion. The model's Hamiltonian is given by $H = \bm{h}(k)\cdot\bm{\sigma}$, with $\sigma$ the Pauli matrix. An exactly flat band structure can be achieved by setting $|\bm{h}|=m$ as a constant. The quantum metric $g$ can be tuned by the normalized vector $\bm{n}(k)=\bm{h}(k)/|\bm{h}|$. We place the model on a lattice with 9 unit-cells (18 sites).
When $h_x = t \cos(ka)$, $h_y = t \sin(ka)$, $g$ is also flat, and no QMO occurs, thus no breathing mode appears, as shown in Figs.~\ref{fig: Oscillations}(a,b). However, if we set $h_x = t \cos(ka)+t' \cos(2ka)$, $h_y = t \sin(ka)+t' \sin(2ka)$, which can be approximated with several short-range hopping terms \footnotemark[\value{footnote}], $g$ fluctuates in momentum space, leading to oscillatory behaviors in the time evolution of the real-space probability density, as illustrated in Figs.~\ref{fig: Oscillations}(c,d). This demonstrates that even in exactly flat bands where BO vanishes, QMO can still survive. We also checked the consistency between numerical simulations and analytical calculations \footnotemark[\value{footnote}].

\textit{Discussion and summary}---
Our proposed QMO is expected to be experimentally observed in various {platforms, especially the} solid superlattices with flat bands that have been used to study BO.
For instance, in 1D epitaxial superlattices where various signals caused by BO have been detected \cite{LEO1992943, PhysRevB.46.7252, PhysRevLett.70.3319, PhysRevB.51.17275, PhysRevLett.64.52, PhysRevLett.64.3167}, flat bands can be engineered %the bands can be designed to be flat, 
such that ZT-induced current is neglectable, BO-induced current dominates at weak fields near half-filling, and QMO-induced current dominates under strong fields when the system is away from breakdown.
2D moir\'e superlattices are also promising platforms, as their tunable band structures and rich quantum geometry would facilitate the observation of both BO- \cite{Bloch_in_Moire} and BCO \cite{Quantum_Geometric_Oscillations, Oscillation_Optical_Effects}-induced drift currents.
We discuss feasible experiments with more details in SM \footnotemark[\value{footnote}].
Moreover, various artificial periodic systems, including optical \cite{PhysRevLett.96.023901, PhysRevLett.102.076802, PhysRevLett.103.123601, PhysRevLett.91.263902, PhysRevLett.83.4756} and cold atom lattices \cite{PhysRevLett.87.140402, PhysRevLett.76.4508}, where BO has been extensively studied, offer potential platforms for observing the oscillatory behaviors of QMO and merit further exploration.
{As a mechanism generally present in periodic systems, QMO opens up new opportunities for studying quantum metrics in these systems.}

In summary, we reveal a strong-field oscillation mechanism induced by interband quantum metric, which can manifest even in exactly flat bands. 
The QMO-induced drift current in a steady state exhibits a linear dependence on the electric field in $\mathcal{T}$- or $\mathcal{P}$-symmetric systems,
%significantly affecting electron dynamics in the strong-field regime 
emerging as the dominant delocalized current in flat bands.
In addition, signals arising from QMO hold promise for detection in systems with flat bands, including 1D epitaxial superlattices and 2D moir\'e lattices.
%Our work not only reveals a new mechanism of strong-field dynamics beyond intraband oscillations and interband tunneling, but also offers a new perspective for studying the quantum metric effcts in periodic systems.

\begin{acknowledgments}
This work is supported by the National Key R\&D Program of China (Grant No. 2021YFA1401600), the National Natural Science Foundation of China (Grants No. 12074006 and 12474056), the Basic Science Center Project of NSFC (Grant No. 52388201), the Ministry of Science and Technology of China, %({To Be Added}), 
and the Beijing Advanced Innovation Center for Future Chip (ICFC). The computational resources were supported by the high-performance computing platform of Peking University.
\end{acknowledgments}

\bibliography{ref}% Produces the bibliography via BibTeX.

\end{document}